\documentclass[9pt,sigconf]{acmart}
\acmConference[  ]{ }{Month DD–DD, 2026}{City, Country}
\acmConference[DAC '26]{Design Automation Conference}{July 26-29, 2026}{Long Beach, CA}
\acmYear{2026}

\def\smallerspacecaption{\vspace{-2mm}}

\usepackage{listings}
\usepackage{xcolor} 
\usepackage{listings}

\usepackage{pifont}
\usepackage{xcolor}

\usepackage{xcolor}

\usepackage{tikz}
\newcommand*\circled[1]{\tikz[baseline=(char.base)]{
            \node[shape=circle,draw,inner sep=1pt] (char) {#1};}}

\usepackage{array,booktabs,colortbl,xcolor}
\definecolor{good}{HTML}{C6E6C1}   
\definecolor{warn}{HTML}{FCE9B2}   
\definecolor{bad}{HTML}{F6C6C6}    
\newcommand{\greencell}{\cellcolor{good}}

\newcommand{\redcell}{\cellcolor{bad}}

\usepackage{array,booktabs,xcolor,listings}

\lstdefinestyle{rtltiny}{
  basicstyle=\scriptsize\ttfamily,
  columns=fullflexible,
  keepspaces=true,
  showstringspaces=false,
  frame=none,
  aboveskip=0pt,
  belowskip=0pt,
  xleftmargin=0pt,
  xrightmargin=0pt
}

\usepackage[most]{tcolorbox}
\usepackage{listings,xcolor,booktabs,array,makecell}
\usepackage{caption}

\tcbset{
  mygreenbox/.style={
    colback=green!10,
    colframe=green!60!black,
    boxrule=0.4pt,
    sharp corners,
    left=1.5pt,
    right=1.5pt,
    top=0pt,        
    bottom=0pt      
  },
  myredbox/.style={
    colback=red!10,
    colframe=red!60!black,
    boxrule=0.4pt,
    sharp corners,
    left=1.5pt,
    right=1.5pt,
    top=0pt,
    bottom=0pt
  }
}

\lstdefinestyle{veritiny}{
  language=Verilog,
  basicstyle=\ttfamily\scriptsize, 
  columns=fullflexible,
  keepspaces=true,
  showstringspaces=false,
  aboveskip=0pt,
  belowskip=0pt
}

\usepackage{xcolor}
\usepackage{mdframed}
\newmdenv[
    frametitlefont=\small\bfseries, 
    frametitlebackgroundcolor=gray!20,
    backgroundcolor=gray!5,
    linecolor=black,
    linewidth=0.8pt, 
    roundcorner=3pt, 
    innertopmargin=0.7\baselineskip, 
    innerbottommargin=0.7\baselineskip, 
    innerleftmargin=6pt, 
    innerrightmargin=6pt,
    skipabove=0.5\baselineskip, 
    skipbelow=0.5\baselineskip 
]{compactpromptbox} 

\usepackage{booktabs,array,xcolor,listings}
\lstdefinestyle{veritiny}{
  language=Verilog,
  basicstyle=\ttfamily\tiny,
  columns=fullflexible,
  breaklines=true,
  keepspaces=true,
  escapeinside={(*@}{@*)} 
}

\usepackage{mdframed}
\lstdefinestyle{svatiny}{
  basicstyle=\ttfamily\scriptsize,
  columns=fullflexible,
  keepspaces=true,
  showstringspaces=false,
  frame=single,
  numbers=left,
  numberstyle=\tiny,
  xleftmargin=1.2em,
  breaklines=true,
  aboveskip=4pt,
  belowskip=4pt,
  escapeinside={(*@}{@*)} 
}

\usepackage{multirow}

\usepackage{amsmath}
\usepackage{bm}
\usepackage{graphicx}
\graphicspath{{incl/}}

\usepackage{booktabs,tabularx,array,multirow}

\usepackage[dvipsnames]{xcolor}
\usepackage{pifont}

\usepackage{listings}
\lstdefinestyle{rtltiny}{
  basicstyle=\scriptsize\ttfamily, columns=fullflexible, keepspaces=true,
  showstringspaces=false, frame=none, aboveskip=0pt, belowskip=0pt,
  xleftmargin=0pt, xrightmargin=0pt
}
\lstdefinestyle{veritiny}{
  language=Verilog, basicstyle=\ttfamily\tiny, columns=fullflexible,
  breaklines=true, keepspaces=true, escapeinside={(*@}{@*)}
}
\lstdefinestyle{svatiny}{
  basicstyle=\ttfamily\scriptsize, columns=fullflexible, keepspaces=true,
  showstringspaces=false, frame=single, numbers=left, numberstyle=\tiny,
  xleftmargin=1.2em, breaklines=true, aboveskip=4pt, belowskip=4pt,
  escapeinside={(*@}{@*)}
}
\lstdefinestyle{svamini}{
  style=svatiny, basicstyle=\ttfamily\scriptsize, aboveskip=2pt, belowskip=2pt,
  framesep=2pt, xleftmargin=0pt, xrightmargin=0pt, breaklines=true, linewidth=\columnwidth
}

\usepackage{subcaption}

\usepackage[ruled,vlined]{algorithm2e}

\usepackage{cleveref}

\usepackage{tikz}

\definecolor{good}{HTML}{C6E6C1}
\definecolor{warn}{HTML}{FCE9B2}
\definecolor{bad}{HTML}{F6C6C6}

\definecolor{SRPurple}{HTML}{6F42C1}

\usepackage{caption}
\lstdefinestyle{svamini}{
  style=svatiny,
  basicstyle=\ttfamily\scriptsize,   
  aboveskip=2pt, belowskip=2pt,      
  columns=fullflexible,
  keepspaces=true,
  framesep=2pt,
  xleftmargin=0pt,
  xrightmargin=0pt,
  breaklines=true, 
  breakatwhitespace=false, 
  linewidth=\columnwidth
}

\usepackage{listings}
\usepackage[most]{tcolorbox} 
\tcbuselibrary{listings,skins,breakable}

\usepackage{enumitem}

\usepackage[dvipsnames]{xcolor}
\definecolor{SRPurple}{HTML}{6F42C1} 

\usepackage{enumitem}
\setlist{nosep,leftmargin=*}

\usepackage{pgfmath}
\usepackage{siunitx}

\newcommand{\timesimp}[2]{%
\pgfmathsetmacro{\factor}{#2/#1}%
\num[round-mode=places,round-precision=2]{#2}%
~(\textcolor{green!50!black}{$\times$\num[round-mode=places,round-precision=2]{\factor}})%
}

\usepackage{fancyhdr}
\fancypagestyle{firstpage}{%
  \fancyhf{}
  
  \fancyhead[C]{\textit{To Appear in 63rd Design Automation Conference (DAC '26), July 26--29, 2026, Long Beach, CA}}
  \fancyfoot[C]{\thepage}
}

\begin{document}

\title{\texttt{STELLAR:} \underline{\textbf{St}}ructur\underline{\textbf{e}}-guided \underline{\textbf{LL}}M \underline{\textbf{A}}ssertion \underline{\textbf{R}}etrieval and Generation for Formal Verification}


\author{Saeid~Rajabi, Chengmo~Yang, and Satwik~Patnaik}
\email{{srajabi, chengmo, satwik}@udel.edu}
\affiliation{
  \institution{Department of Electrical \& Computer Engineering, University of Delaware, Newark, DE, US}
  \country{}
}

\begin{abstract}
Formal Verification (FV) relies on high-quality SystemVerilog Assertions (SVAs), but the manual writing process is slow and error-prone.
Existing LLM-based approaches either generate assertions from scratch or ignore structural patterns in hardware designs and expert-crafted assertions.
This paper presents \texttt{STELLAR}, the first framework that guides LLM-based SVA generation with structural similarity. \texttt{STELLAR} represents RTL blocks as AST structural fingerprints, retrieves structurally relevant (RTL, SVA) pairs from a knowledge base, and integrates them into structure-guided prompts.
Experiments show that \texttt{STELLAR} achieves superior syntax correctness, stylistic alignment, and functional correctness, highlighting structure-aware retrieval as a promising direction for 
industrial FV.

\end{abstract}


\maketitle
\thispagestyle{firstpage}
\section{Introduction}
\label{sec:introduction}

Formal Verification (FV) is vital in the VLSI design flow, addressing the growing complexity
of hardware architectures~\cite{gupta1992formal}. In particular, 
FV applies mathematical methods to prove that a design satisfies a given property, offering stronger guarantees than simulation. 
A widely used approach is Assertion-Based Verification (ABV)~\cite{foster2008assertion,witharana2022survey}, which relies on SystemVerilog Assertions (SVAs) to formally specify the expected behavior of a design. 
SVAs can be verified either statically with formal tools or dynamically in simulation.
The effectiveness of FV depends heavily on the quality of SVAs~\cite{witharana2022survey}. 
However, manually writing high-quality SVAs is time-consuming, error-prone, and difficult to scale, as it requires translating often ambiguous natural language specifications (SPEC) into precise formal properties. 
Automating assertion generation is crucial for fully realizing the potential of FV in modern hardware verification.
The 
main challenge
lies in bridging the semantic gap between the high-level design intent, 
which is often described ambiguously in SPEC, and the low-level implementation details 
in the Register-Transfer Level (RTL) code. 
Verification engineers must mentally synthesize information from both sources to create high-quality assertions, a process that 
is
challenging
to automate.

\begin{figure}[t]
    \centering
    \includegraphics[width=0.97\linewidth]{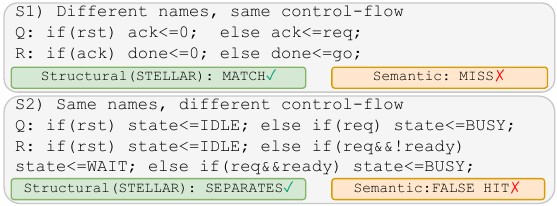}
    \smallerspacecaption
    \smallerspacecaption
    \caption{Structural vs.\ semantic retrieval: S1 (different names, same control-flow) structure-aware correctly matches but semantic may miss; S2 (different control-flow, same name) semantic has a false hit while structure-aware separates.}
    \label{fig:struct-semantic-2boxes}
\end{figure}

Despite the capabilities of large language models (LLMs) in code generation, a 
recent study indicates that commercial LLMs alone are not yet 
fully equipped
for assertion generation, as they often produce syntactic or semantic errors~\cite{pulavarthi2025llms}.
Other recent work has shown potential improvement for LLM-based automating assertion generation. 
For example, SANGAM~\cite{gupta2025sangam} performs iterative reasoning over multimodal specifications and RTL, while AssertionForge~\cite{bai2025assertionforge} constructs structured knowledge from SPEC and RTL to synthesize comprehensive design information.
Despite these advancements, both works share a \textbf{key limitation} as they aim to create assertions from scratch by only analyzing the underlying design files (SPEC and RTL), without the benefit of learning from existing expert-written assertions in industry codebases.
For instance, AssertionForge 
provides limited context to the LLM regarding the information it extracts from the current design's SPEC and RTL, including code snippets and information based on matching signal names. 
This 
``invent from scratch'' approach overlooks the most valuable asset in the industrial FV setting, which is codebases of expert-written RTL and SVA pairs from past/existing projects. 
Failing to utilize these proven examples results in several limitations including wasted efforts, 
the reintroduction of 
previously resolved bugs, and the failure to enforce the consistent, industry-standard SVA styles found in expert-written codebases.

To bridge this gap, this work proposes \texttt{STELLAR} \textit{(\underline{St}ructur\underline{e}-guided \underline{LL}M \underline{A}ssertion \underline{R}etrieval and Generation for Formal Verification)}, the first framework built upon the hypothesis that structural similarity serves as a more effective metric for retrieval than semantic similarity for SVA generation. 
Since assertions are inherently tied to the structural patterns of the design, retrieval methods that overlook this aspect miss a powerful signal for guiding LLMs.
As illustrated in Fig.~\ref{fig:struct-semantic-2boxes}, which shows that structure-aware retrieval can correctly \emph{match} when control flow aligns despite different naming (S1) and can \emph{separate} when names align but control flow diverges (S2).

Instead of inventing assertions or using flawed semantic retrieval, \texttt{STELLAR} intelligently reuses an existing knowledge base of (RTL, SVA) pairs by identifying similar structural patterns. 
These relevant, high-quality examples from the knowledge base are used to guide the LLM to generate accurate and stylistically consistent SVAs.
The primary technical contributions include:

\begin{itemize}[align=parleft,leftmargin=*,noitemsep,nolistsep]
\item \textbf{\texttt{STELLAR} Framework,} the first structure-aware retrieval pipeline leveraging structural fingerprints for 
    efficient retrieval of 
    structurally similar (RTL, SVA) 
    pairs from the knowledge base; 

\item \textbf{Structural Fingerprinting:} a novel AST-based\footnote{Abstract Syntax Tree (AST) is a tree-based representation of code structure capturing operations and control-flow.} 
    fingerprinting technique encoding control flow, complexity, and context of RTL blocks for fine-grained structural comparison.
    
\item \textbf{Structure-Guided Prompting:} an effective prompting strategy utilizing retrieved context and a dynamically computed execution path count to ensure
    a high-coverage, logically consistent set of 
    SVA generation by LLM;
  
\item \textbf{Comprehensive Evaluation} of \texttt{STELLAR} against multiple baselines,  demonstrating its superiority in (i)~syntactic correctness, (ii)~semantic alignment, and (iii)~functional correctness of the generated assertions.
\end{itemize}

The rest of this paper is organized as follows. Section~\ref{sec:related_work} reviews related work. Section~\ref{sec:STELLAR} details the \texttt{STELLAR} framework. Section~\ref{sec:exp_setup} describes our experimental setup, and Section~\ref{sec:results} presents our results. Finally, Section~\ref{sec:Conclusion} concludes the paper.
\section{Background and Related Works}
\label{sec:related_work}

\subsection{Automated Assertion Generation}
The manual effort and expertise required for high-quality assertion generation have motivated the research into automating SVA 
generation~\cite{kande2024security,kang2025fveval,liu2024domain,orenes2021autosva}.
Non-LLM methods such as GOLDMINE~\cite{vasudevan2010goldmine}, A-Team~\cite{danese2017team}, and HARM~\cite{germiniani2022harm}
rely on data mining, templates, and/or static analysis. 
While effective in certain domains, they often fail to scale to large industrial designs due to the algorithmic complexity of static analysis~\cite{pulavarthi2025llms} and state explosion, which limits their capability on large designs~\cite{witharana2022survey}.

LLM-based approaches have gained a lot of research attention recently~\cite{sun2023towards,maddala2024laag,menon2025openassert,ankireddy2025lasa}.
ChIRAAG~\cite{mali2024chiraag} automates assertion generation from natural language SPECs using GPT-4 with iterative refinement.
Likewise, AssertLLM~\cite{yan2025assertllm} uses multiple specialized LLMs to extract information from SPECs and generate SVAs, with RTL only used for verification.
SANGAM~\cite{gupta2025sangam} uses multi-modal SPEC files, including diagrams, with Monte Carlo Tree Search for iterative refinement for SVA generation. 
SANGAM~\cite{gupta2025sangam} uses RTL 
for signal-name mapping between RTL and SPEC and final verification, whereas AssertionForge~\cite{bai2025assertionforge} combines SPEC files and RTL into a unified Knowledge Graph to guide the LLM  
for SVA
generation.\footnote{Knowledge Graph is a graph-structured representation of entities (nodes) and relations (edges).}
A critical limitation of 
prior LLM-based methods is that they generate SVAs from scratch or use context limited to the target design's SPECs and RTL. 
\subsection{Retrieval-Augmented Generation (RAG)}

LLMs are powerful but prone to hallucination when not grounded in a reliable context.
RAG~\cite{lewis2020retrieval}, originating in natural language processing tasks, mitigates hallucination by retrieving relevant context from an external knowledge base so that context generation is guided by 
prior knowledge rather than the model's world knowledge.
RAG uses embedding models to convert text or code into dense vector representations that capture semantic meaning. 
At retrieval time, RAG finds the items whose embeddings are closest to the query (\textit{e.g.,} by cosine similarity), 
providing semantically related context to the LLM even when the wording differs.
\subsection{LLM Interaction Paradigms}
Table~\ref{tab:llm-paradigms} shows the different modes that LLMs can be used.
The \textbf{zero-shot} setting 
relies solely on pretrained knowledge.
\textbf{RAG}, rather than relying solely on pretraining, improves reliability by grounding the model with relevant context retrieved from an external knowledge base.
\textbf{Fine-tuning} adapts model weights using task-specific data, which can improve performance but (i) requires large, curated labeled datasets; (ii) is computationally expensive; (iii) may cause catastrophic forgetting when learning new information;
and (iv) can reduce cross-task generalizability.
In hardware verification, RAG emerged recently. AssertionForge~\cite{bai2025assertionforge} retrieves context via semantic similarity from RTL and SPEC. However, as shown in Fig.~\ref{fig:struct-semantic-2boxes}, semantic similarity alone often miss the structural control flow essential for assertions, motivating structure-aware retrieval.

\begin{table}[t]
\centering
\smallerspacecaption
\caption{Comparison of LLM interaction paradigms}
\label{tab:llm-paradigms}
\setlength{\tabcolsep}{4pt} 
\renewcommand{\arraystretch}{0.95}
\begin{tabular}{@{}lccc@{}}
\toprule
\textbf{Feature} & \textbf{Zero-shot} & \textbf{Fine-tune} & \textbf{RAG (Ours)} \\ 
\midrule
Training Required      & \greencell No  & \redcell Yes   & \greencell No \\ 
Retraining (New Data)  & \greencell No  & \redcell Yes   & \greencell No \\ 
Hallucination Risk     & \redcell High  & \greencell Low & \greencell Low \\ 
Generalizability       & \greencell High & \redcell Low   & \greencell High \\ 
\bottomrule
\end{tabular}
\footnotesize 
\textit{Color legend:} 
\colorbox{good}{\strut good},\;
\colorbox{bad}{\strut poor}.
\end{table}

\subsection{Hardware Similarity and Structural Representations}
\label{sec:hw_similarity}
Measuring hardware similarity
has been explored at gate level~\cite{li2022deepgate} and RTL~\cite{yu2021hw2vec,akyash2025simeval}. HW2VEC~\cite{yu2021hw2vec}
converts RTL into graph 
and uses GNNs to generate embeddings.
While effective for classification tasks (\textit{e.g.,} detecting Trojans), it lacks the fine-grained structural resolution needed for SVA retrieval.
SimEval~\cite{akyash2025simeval} evaluates LLM-generated RTL code using 
syntax, 
control-flow, and functionality, but oversimplifies results into a single scalar score. 
If multiple candidates receive same scores for a given query, SimEval provides no
tie-break, rendering distinct designs indistinguishable.

Overall, HW2VEC shows the power of learned embeddings but lacks fine-grained retrieval resolution, while SimEval offers multi-level similarity but no retrieval capability.
For SVA generation, these works underscore the need for a structural-aware retrieval approach that distinguishes nuanced RTL patterns and provides assertion-aligned context.

\subsection{Dataset Generation for Assertion Learning}

A major challenge in automating assertion generation is the lack of large, standardized datasets,
as open-source projects often use inconsistent assertion styles~\cite{menon2025vert}. 
Nonetheless, hardware designs are typically verified at the module level, which aligns well with the context limits of LLMs.
VERT~\cite{menon2025vert} addresses this gap by providing the first large-scale benchmark of 20,000 paired RTL code and SVAs, drawn from open-source System-on-Chip designs \cite{zhao2020sonicboom, lee2016rocketchip, xu2022xiangshan}, with augmentation to diversify variable names and control structures. 
VERT systematically includes patterns such as deeply nested \textit{if–else} statements, \textit{synchronous/asynchronous} blocks, where LLMs often 
fails~\cite{menon2025vert}.
\texttt{STELLAR} uses VERT as the knowledge base, enabling scalable structure-aware retrieval of similar (RTL, SVA) pairs.

\section{Methodology: \texttt{STELLAR} Framework}
\label{sec:STELLAR}
\texttt{STELLAR} generates SVAs by retrieving structurally matched RTL and guiding an LLM through a four-stage pipeline:
(i) structural fingerprinting, (ii) retrieval process, (iii) structural-aware prompting, and (iv) SVA generation.
Each stage addresses key challenges through targeted solutions.
To ensure scalability, \texttt{STELLAR} first decomposes each RTL module into individual \textit{always} blocks or FSM units paired with the essential local context (signal declarations, sensitivity lists, and interface signals) involved in cross-block behavior, so that SVA generation remains accurate even if the entire RTL does not fit into a single prompt. This block-level formulation preserves protocol semantics while enabling efficient SVA retrieval and generation across large designs.

\subsection{Structural Fingerprinting}
\label{sec:structural_fingerprinting}

Assertion logic is closely linked to the hardware's structure, and even minor changes in control/data-flow require different SVAs. 
Thus, an effective automated approach must identify key RTL structures to find relevant examples to guide an LLM for SVA generation.

\begin{figure}[ht]
    \centering
    \includegraphics[width=0.985\linewidth]{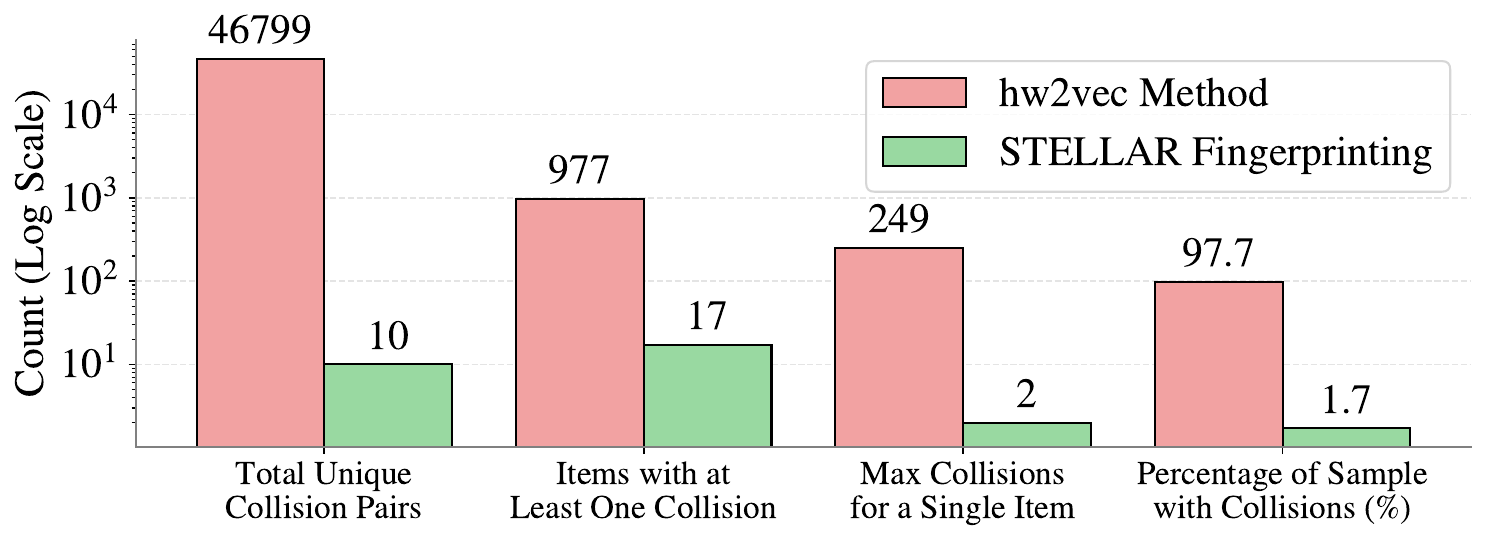}
    \smallerspacecaption
    \smallerspacecaption
    \caption{Collision comparison for HW2VEC~\cite{yu2021hw2vec} embeddings and \texttt{STELLAR} structural fingerprinting (lesser is better).}
    \label{fig:comparison_clara_hw2vec}
\end{figure}

\indent\textbf{Challenge~\circled{1} 
Semantic vs.\ Structural Similarity.} The primary challenge in retrieving relevant examples for SVA generation is choosing the correct similarity metric. 
\textbf{Semantic similarity}, common in RAG, depends on textual content and variable names but can be misleading in hardware design due to arbitrary signal names 
that do not reflect logic.
Relying solely on semantics leads to retrieving structurally divergent code that happens to share variable names (as shown in Fig.~\ref{fig:struct-semantic-2boxes}), providing misleading context to the LLM.
In comparison, \textbf{structural similarity} is a more reliable indicator. 
However, not all methodologies are effective (Sec.~\ref{sec:hw_similarity}). 
GNN-based models such as HW2VEC~\cite{yu2021hw2vec} produce coarse embeddings, resulting in vector collapse and 
collisions, 
whereby distinct RTL blocks whose embeddings have perfect cosine similarity are mapped to identical vectors. 
To quantify this, we performed experiments to analyze the embedding space of HW2VEC on 1K randomly selected designs from the VERT dataset. 
The result is shown in Fig.~\ref{fig:comparison_clara_hw2vec}. 
The fact that 97.7\% of the designs were involved in at least one collision confirms that HW2VEC embeddings are unsuitable for distinguishing logic and accurate retrieval.

\indent\textbf{Solution~\circled{1}.} To capture structural fingerprinting, \texttt{STELLAR} employs a deterministic AST-based signature. 
It generates a hierarchical string that encodes key structural features, including control-flow patterns (\textit{e.g.,} if-else nesting, depth, and branching count), hardware description language assignment types (\textit{e.g.,} blocking = vs.\ non-blocking $<=$), and data-flow logic. 
This 
approach enables high-resolution representations, reducing the collision rate from 97.7\% (HW2VEC) to 1.7\% (Fig.~\ref{fig:comparison_clara_hw2vec}).
Our results demonstrate that a high-granularity structural representation is essential to 
distinguish RTLs
and is insensitive to declaration distance or glue logic.

\textbf{Challenge~\circled{2} Lack of Contextual Distinction.} Generic structural representations often fail to capture critical hardware context
or distinguish combinational from sequential logic or different clock sensitivities (\textit{e.g.,} \textit{posedge} vs.\ \textit{negedge}).
A retrieval method that confuses these contexts will retrieve flawed examples, leading to incorrect SVA generation.

\textbf{Solution~\circled{2}.} To address this challenge, \texttt{STELLAR}'s fingerprinting is context-aware by design. Specifically, it tags each fingerprint with a context-defining prefix (\textit{e.g.}, \textit{ASYNC::} or \textit{SYNC\_POSEDGE::}) based on the \emph{always} block's sensitivity list. With such information, when the RAG pipeline retrieves examples for a sequential query, it correctly deprioritizes combinational logic with a similar control-flow, leading to more relevant and contextually appropriate results. Table~\ref{tab:rtl_clara_eg} shows an example of \texttt{STELLAR} structural fingerprinting.

\begin{table}[ht]
\centering
\scriptsize
\setlength{\tabcolsep}{2pt}    
\renewcommand{\arraystretch}{0.7} 
\begin{tabular}{@{}p{0.5\linewidth}p{0.48\linewidth}@{}} 
\toprule
\textbf{\vspace{-2pt}RTL Snippet} & \textbf{\vspace{1pt}Simplified \texttt{STELLAR} Fingerprint} \\[-1pt] 
\midrule
\begin{minipage}[t]{\linewidth}
\vspace{-2pt}
\begin{lstlisting}[style=rtltiny]
always @(posedge clk) begin
    if (a && b) begin
      out <= c + 1;
    end else ...
\end{lstlisting}
\end{minipage}
&
\begin{minipage}[t]{\linewidth}
\raggedright

\textcolor{orange}{\texttt{SYNC POSEDGE::}}\\
\textcolor{blue}{\texttt{if(dpth:1,brnch:2,...)}}\\
\textcolor{red}{\texttt{block(nb\_asgn:1,...)}}\\
\textcolor{green!50!black}{\texttt{-else(block(b\_asgn:1,...))}}
\end{minipage} \\
\bottomrule
\end{tabular}
\vspace{2pt}
\caption{Example of \texttt{STELLAR} structural fingerprinting}
\label{tab:rtl_clara_eg}
\end{table}

\vspace{-10pt}
\begin{figure}[ht!]
    \centering
    \vspace{-9pt}
    \includegraphics[width=0.94\linewidth]{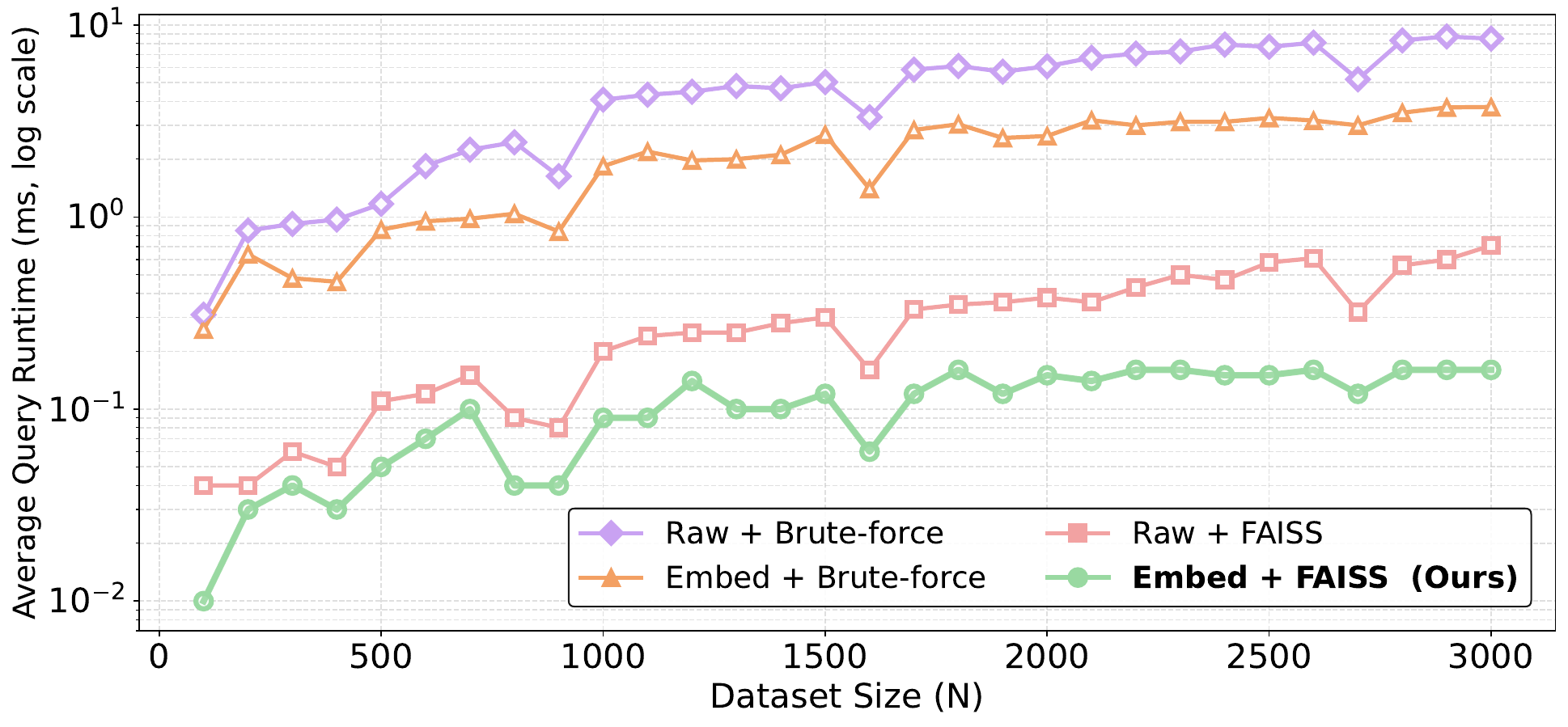}
    \smallerspacecaption
    \smallerspacecaption
    \vspace{-3pt}
    \caption{Query runtime for raw string match (RAW) vs.\ embedding retrieval using brute-force and FAISS indexing.}
    \label{fig:ablation_runtime}
\end{figure}

\begin{figure}[t]
    \centering
    \includegraphics[width=0.88\linewidth]{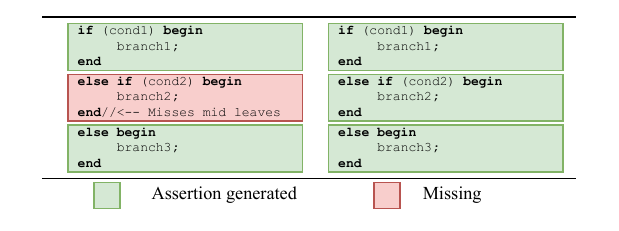}
    \smallerspacecaption
    \smallerspacecaption
    \vspace{-2pt}
    \caption{\footnotesize Structure-guided prompting: left = without, right = with.}
    \label{fig:structure_prompting_clara_eg}
    \includegraphics[width=0.94\linewidth]{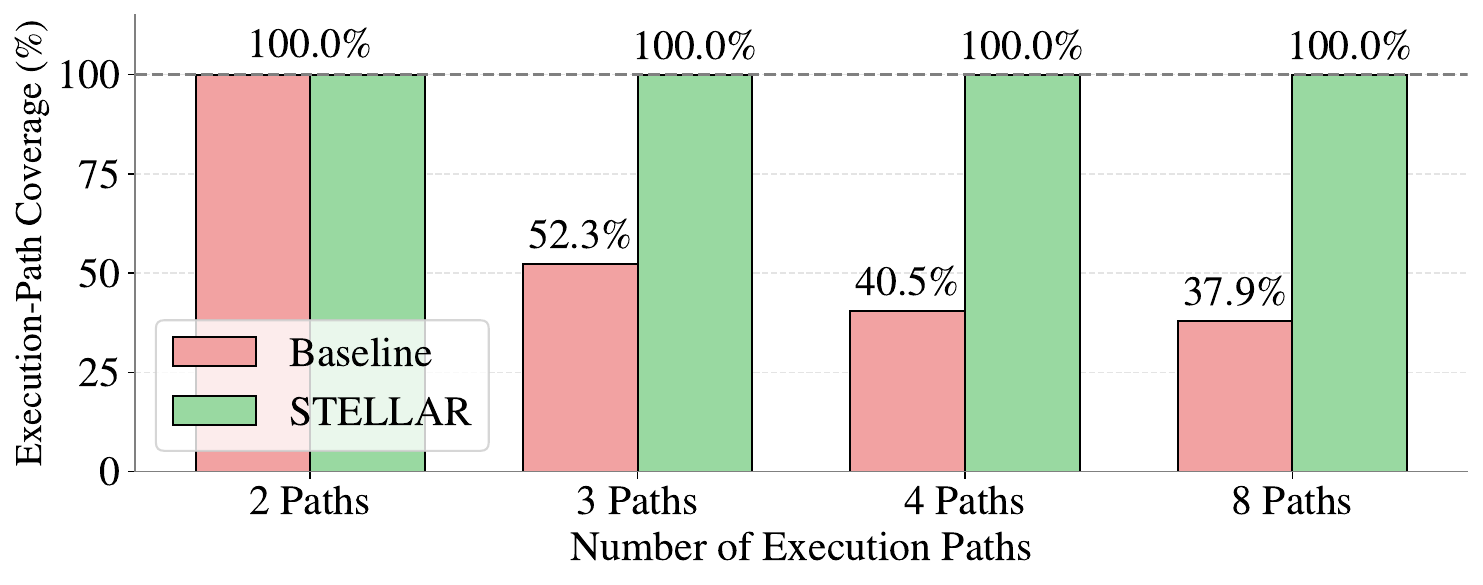}
    \smallerspacecaption
    \vspace{-2pt}
    \caption{Execution-path coverage (higher is better)}
    \label{fig:Precision_path_count_clara}
\end{figure}

\begin{figure*}[ht]
    \centering
\includegraphics[width=0.99\textwidth,height=0.254\textheight,keepaspectratio]
    {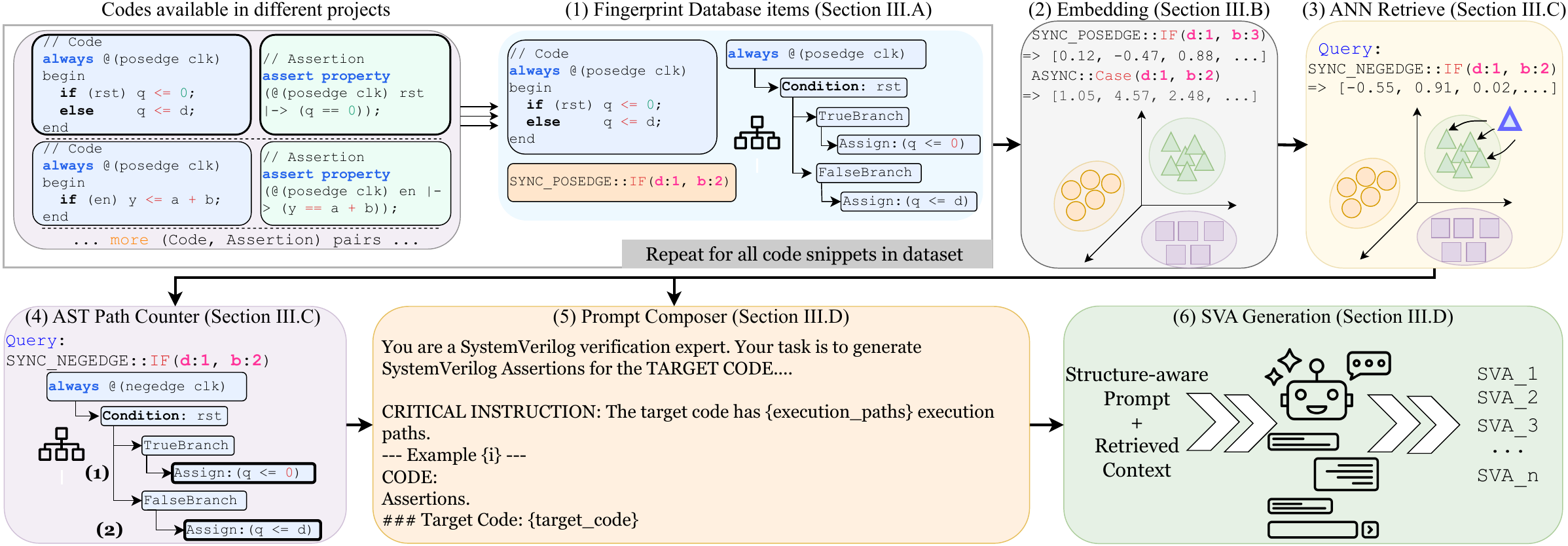}
    \smallerspacecaption
    \captionof{figure}{Architecture of \texttt{STELLAR} framework}
    \label{Fig:CARAG_framework_v1}
\end{figure*}

\subsection{Structural Embeddings and Retrieval}
\textbf{Challenge~\circled{3} Scalability of Retrieval.} While \texttt{STELLAR}'s fingerprints are distinctive, their reliance on string-based data creates a retrieval challenge. 
Character-by-character string matching is not only computationally expensive
but also overly rigid such that minor, logically irrelevant reordering of statements can produce a different string, preventing a correct match.

\textbf{Solution~\circled{3}.} To enable efficient
similarity search, the structural fingerprints are converted into dense vector embeddings using the lightweight, efficient \texttt{all-MiniLM-L6-v2} Sentence Transformer~\cite{allminilm-l6-v2}, suitable for encoding short, structured texts into a semantic space where proximity indicates similarity.

Furthermore, instead of performing vector search which is too slow at scale, \texttt{STELLAR} employs FAISS (Facebook AI Similarity Search). 
This library creates an index that facilitates Approximate Nearest Neighbor (ANN) search, allowing quick navigation to likely candidates instead of exhaustive comparisons. 
As shown in Fig.~\ref{fig:ablation_runtime}, FAISS-based embedding retrieval achieves significant speedup in the RAG pipeline as the dataset size grows, compared to raw string matching and brute-force search.  
These data are collected by populating the dataset with every 100 samples up to 3,000 samples 
and reporting the query runtime for each setting.

\subsection{Structure-Guided Prompting}

\textbf{Challenge~\circled{4} Incomplete Execution-Path Coverage.} A primary failure mode for LLMs in SVA generation, even with RAG, is missing required execution paths. As probabilistic models, LLMs do not deterministically trace every logical path within an RTL block to ensure full coverage. 
Therefore, for an \textit{if-else} chain (Fig.~\ref{fig:structure_prompting_clara_eg}), an LLM might generate properties for 
the 
initial if and the final else branch, while completely ignoring 
the 
intermediate conditions. 
This creates critical blind spots in the verification, leaving parts of the design unverified. 

\textbf{Solution~\circled{4}.} \texttt{STELLAR} addresses the aforementioned challenge with a structure-guided prompting strategy enforcing 
execution-path coverage.
Before querying LLM, \texttt{STELLAR} dynamically traverses the AST of the RTL to 
count
unique execution paths. 
This count is injected into prompt as a CRITICAL INSTRUCTION:
\begin{compactpromptbox}\small 
\texttt{\textbf{CRITICAL INSTRUCTION}}: Target code has \{\#exec\_path\} execution paths. You must generate exactly \{\#exec\_path\} assertions, one for each path.
\end{compactpromptbox}

Fig. 5 quantifies the effectiveness of \texttt{STELLAR}'s structure-guided prompting. Zero-shot baseline (\texttt{GPT-4o mini} with no examples) shows a downward trend as design complexity increases, with execution-path coverage dropping from 100\% in 2-path designs to 38\% in 8-path designs. 
In contrast, \texttt{STELLAR} maintains near-100\% path coverage across all complexities, underscoring the importance of structure-aware prompting for generating high-coverage SVAs.

\subsection{SVA Generation Using LLMs}

\texttt{STELLAR} integrates the top-\textit{k} retrieved, structurally similar pairs (RTL, SVA) and the structure-guided CRITICAL INSTRUCTION constraint with the target RTL to construct a comprehensive prompt. 
This complete context enables the LLM to synthesize the structural patterns and assertion styles from the retrieved context, guided by the explicit execution path count. 
This guides the LLM toward generating a high-coverage and stylistically consistent set of SVAs. 
The model's raw output is then parsed to extract the final assertion, ready for verification.
\subsection{Putting It All Together}

\texttt{STELLAR} operates in two phases: an offline phase for knowledge base construction and an online phase for SVA generation (Fig.~\ref{Fig:CARAG_framework_v1}).

\begin{enumerate}[align=parleft,leftmargin=*,noitemsep,nolistsep]

\item \textbf{Offline Phase: Knowledge Base Construction.} In industrial settings, verification teams have access to vast repositories of existing RTL blocks, each paired with expert-written SVAs. 
\texttt{STELLAR} leverages this valuable codebase by first constructing a reusable, structure-aware knowledge base in an offline phase. For each valid RTL and corresponding SVA,  its AST is parsed to generate a structural fingerprint with its hardware context tags (Step~\circled{1}).
Then, all fingerprints are converted into vector embeddings (Step~\circled{2}) and stored in a FAISS index for scalable retrieval to perform ANN search.

\item \textbf{Online Phase: SVA Generation.} Given a target RTL block, \texttt{STELLAR} parses its AST to generate a unique fingerprint. Then, based on its embedding, \texttt{STELLAR} retrieves the top-\textit{k} most structurally similar (RTL, SVA) examples based on the ANN search (Step~\circled{3}). 
It calculates the execution path count of the target code (Step~\circled{4}) and assembles retrieved examples, target RTL, and CRITICAL INSTRUCTION into a structure-guided prompt (Step~\circled{5}) to generate assertions.

\end{enumerate}
\section{Experimental Setup}
\label{sec:exp_setup}

\subsection{Implementation Details}

\texttt{STELLAR} uses PyVerilog~\cite{Takamaeda2015Pyverilog} 
for AST parsing. 
LLM temperature is set to 0 
for deterministic outputs
, and 
a 1024-token limit.
Performance of \texttt{STELLAR} is evaluated by incorporating the top-3/top-5 most structurally similar examples into the structure-aware prompt, along with target RTL code.

\subsection{Knowledge Base Curation and Stratification}
Our experiments leverage the latest version of VERT dataset~\cite{menon2025vert} as the foundational source.
A preliminary analysis revealed that a subset of these pairs contained syntactic errors.
The following examples show two such cases that were filtered out: one with an extra parenthesis and another with an invalid event control placement within a Boolean expression.

\begin{lstlisting}[style=svatiny,language=Verilog,basicstyle=\scriptsize\ttfamily,
  caption={Invalid SVA: extra ')' (VERT \#39)},
  label={lst:vert39-invalid},aboveskip=3pt,belowskip=3pt]
property RXSynceotid;
  (interrupt_control_14) != 7'b0110x11) (*@\textcolor{red}{)}@*)) |=> rx_14 == core_1;
endproperty
\end{lstlisting}
\begin{lstlisting}[style=svatiny,language=Verilog,basicstyle=\scriptsize\ttfamily,
  caption={Invalid SVA: clock in antecedent (VERT \#17)},
  label={lst:vert17-mid-event},aboveskip=3pt,belowskip=3pt]
property SyncReseteotid;
  (status_register_status_10) != 6'bxx0x0x &&
  (status_register_status_10) != 7'b00x0001 &&
  (*@\textcolor{red}{@(negedge fast\_clk\_8)}@*) (status_register_status_10) != 7'b1x1xxxx
  |-> hw_4 == rx_5 && cfg_11 == sig_17;
endproperty
\end{lstlisting}

Using this raw, noisy data would directly compromise the RAG's effectiveness, degrading LLM's performance and generating incorrect SVA.
To ensure knowledge base integrity,
we implemented a crucial pre-processing step before indexing which performed (i) SystemVerilog syntax check on every SVA, (ii) discarding failed entries, and (iii) normalizing formatting. 
The cleaned dataset is then split into a knowledge base and a separate, non-overlapping query set. 
To ensure a representative distribution of diverse hardware patterns in the knowledge base and the test set, the cleaned dataset is arranged based on three key structural features: (i) the number of execution paths, (ii) the timing behavior (synchronous vs.\ asynchronous), and (iii) the primary control structure (\textit{e.g.,} if-else vs.\ case). 
The data was split using a 2:1 ratio, resulting in around 10,000 elements for the knowledge base and 5,000 for the query set.
\subsection{Models and Baselines}
To showcase the 
applicability  of \texttt{STELLAR}, it is 
integrated
into
3
LLM classes to evaluate its effectiveness: (i) proprietary (\texttt{GPT-4o mini}), (ii) general-purpose open-source (\texttt{Llama-3.1- 8b-Instruct}), and (iii) domain-adapted to coding (\texttt{CodeLlama- 7b-Instruct}).
\subsection{Evaluation Metrics}

We evaluate generated assertions on a stratified query set, comparing with and without \texttt{STELLAR} across three LLMs using top-3 and top-5 retrieved examples. Performance is measured by three metrics: (i) syntax correctness, 
(ii) similar to~\cite{kang2025fveval}, BLEU score for lexical similarity to golden SVA and BERT score for semantic similarity 
to the reference
SVAs\footnote{Since VERT provides a canonical reference assertion for each RTL snippet, these similarity metrics serve as lightweight proxies for functional alignment and correctness relative to the intended behavior encoded with respect to that reference.},
(iii) the execution paths covered by the generated assertions for each RTL block.

\section{Results}
\label{sec:results}

\begin{table*}[ht]
\centering
\smallerspacecaption
\smallerspacecaption
\caption{
Assertion generation quality for STELLAR-guided and zero-shot LLMs. 
\textit{K} is the number of retrieved structurally relevant examples included in the prompt. BLEU and BERTScore are in [0,1]; parentheses denote improvement over baseline.}
\smallerspacecaption
\renewcommand{\arraystretch}{0.8}
\label{tab:end_to_end_quality}
\scriptsize
\begin{tabular}{@{}l|ccc|ccc|ccc@{}}
\toprule
\multirow{2}{*}{\textbf{Metric}}

  & \multicolumn{3}{c|}{\textbf{\texttt{Llama-3.1-8b}}}
  & \multicolumn{3}{c|}{\textbf{\texttt{CodeLlama-7b}}}
  & \multicolumn{3}{c}{\textbf{\texttt{GPT-4o mini}}} \\
\cmidrule(l{10pt}r{10pt}){2-4}\cmidrule(l{10pt}r{10pt}){5-7}\cmidrule(l{10pt}r{10pt}){8-10}
& \textbf{Zero-Shot} & \textbf{\texttt{STELLAR} (\textit{K}=3)} & \textbf{\texttt{STELLAR} (\textit{K}=5)}
& \textbf{Zero-Shot} & \textbf{\texttt{STELLAR} (\textit{K}=3)} & \textbf{\texttt{STELLAR} (\textit{K}=5)}
& \textbf{Zero-Shot} & \textbf{\texttt{STELLAR} (\textit{K}=3)} & \textbf{\texttt{STELLAR} (\textit{K}=5)} \\
\midrule
\textbf{Syntax Correctness (SC) (\%)} 
& 33.25 & \timesimp{33.25}{61.37} & \textbf{\timesimp{33.25}{77.29}}
& 46.26 & \timesimp{46.26}{78.63} & \textbf{\timesimp{46.26}{84.44}}
& 84.88 & \timesimp{84.88}{99.20} & \textbf{\timesimp{84.88}{99.80}} \\
\midrule
\textbf{Functional Similarity (BLEU)} 
& 0.33  & \timesimp{0.33}{0.80}   & \textbf{\timesimp{0.33}{0.88}}
& 0.255 & \timesimp{0.255}{0.83}  & \textbf{\timesimp{0.255}{0.836}}
& 0.56  & \timesimp{0.56}{0.929}  & \textbf{\timesimp{0.56}{0.932}} \\
\textbf{Semantic Similarity (BERTScore F1)}
& 0.876 & \timesimp{0.876}{0.944} & \textbf{\timesimp{0.876}{0.952}}
& 0.874 & \timesimp{0.874}{0.937} & \textbf{\timesimp{0.874}{0.948}}
& 0.904 & \timesimp{0.904}{0.954} & \textbf{\timesimp{0.904}{0.957}} \\
\midrule
\textbf{Execution-Path Coverage (\%)}
& 29.52 & \timesimp{29.52}{65.53} & \textbf{\timesimp{29.52}{93.04}}
& 29.49 & \timesimp{29.49}{72.13} & \textbf{\timesimp{29.49}{73.38}}
& 49.84 & \timesimp{49.84}{100.0} & \textbf{\timesimp{49.84}{99.92}} \\
\bottomrule
\end{tabular}
\end{table*}

This section evaluates \texttt{STELLAR} by first demonstrating RAG retrieval quality, then SVA generation performance, and finally UART, I\textsuperscript{2}C, and SHA-3 case studies.%

\subsection{Retrieval Quality}

Since the VERT dataset~\cite{menon2025vert} lacks a ground truth for RTL similarity, our retrieval mechanism is evaluated with a controlled synthetic experiment. Specifically, we select 100 random RTL modules from the knowledge base for duplication and run two cases: (1) exact duplication (0\% renaming), and (2) 
identifier names are changed (30\% renaming) while preserving functionality.
For each case, we query the retrieval system to assess whether it can rank the original source high.
The results are shown in Table \ref{tab:robustness_alpha}, with \textit{N} denoting the top-\textit{N} retrieved results, \textit{Recall@N} measuring the success rate of retrieving the correct item within the top \textit{N} results, 
\textit{MRR@N} reporting how early the first correct item appears, and \textit{nDCG@N} measuring ranking quality.
All values are normalized to the range of [0, 1]. The higher, the better.

As shown in Table \ref{tab:robustness_alpha},
when no renaming is applied,
the baseline that relies on semantic embedding of raw RTL code
perfectly retrieves the matching source.
However, its performance degrades significantly (\textit{Recall@10} drops from 1.0 to 0.2) when 30\% of the variables are renamed. This is because the baseline, which directly encodes raw RTL using \texttt{all-MiniLM-L6-v2}~\cite{allminilm-l6-v2} without fingerprinting, is highly sensitive to the change in code lexical representation.
In contrast, \texttt{STELLAR} uses structural fingerprinting invariant to variable names, allowing it to remain stable under both settings.
These results also show that \texttt{STELLAR} retrieves and ranks relevant RTL codes high even when substantial naming differences are introduced, highlighting its superior practicality to baseline.

\begin{table}[ht]
\centering
\smallerspacecaption
\caption{Semantic (Base) vs.\ Structural (STELLAR) retrieval quality under 
1) exact duplication and 2) 30\% variable renaming. Data are computed over top-\textit{N} retrieved results and normalized to [0, 1].}
\renewcommand{\arraystretch}{0.8}
\scriptsize
\begin{tabular}{@{}l
                cc @{\hspace{8pt}}
                cc @{\hspace{8pt}}
                cc @{\hspace{8pt}}
                cc@{}}
\toprule
\multirow{2}{*}{\textbf{Metric}} 
  & \multicolumn{2}{c}{\textbf{N=3}}
  & \multicolumn{2}{c}{\textbf{N=5}}
  & \multicolumn{2}{c}{\textbf{N=7}}
  & \multicolumn{2}{c}{\textbf{N=10}} \\
\cmidrule(lr){2-3}\cmidrule(lr){4-5}\cmidrule(lr){6-7}\cmidrule(lr){8-9}
  & \textbf{Base} & \textbf{\texttt{STELLAR}}
  & \textbf{Base} & \textbf{\texttt{STELLAR}}
  & \textbf{Base} & \textbf{\texttt{STELLAR}}
  & \textbf{Base} & \textbf{\texttt{STELLAR}} \\
\midrule
\multicolumn{9}{c}{\textbf{No variable renaming (Exact duplication)}} \\ 
\midrule
\textbf{Recall@N} 
 & 1.000 & 0.96
 & 1.000 & 0.98
 & 1.000 & 0.98
 & 1.000 & 0.99 \\
\textbf{MRR@N} 
 & 1.000 & 0.472
 & 1.000 & 0.476
 & 1.000 & 0.476
 & 1.000 & 0.477 \\
\textbf{nDCG@N} 
 & 1.000 & 0.599
 & 1.000 & 0.607
 & 1.000 & 0.607
 & 1.000 & 0.610 \\
\midrule
\multicolumn{9}{c}{\textbf{30\% variable renaming (Same logic, different variable names)}} \\ 
\midrule
\textbf{Recall@N} 
 & 0.18 & 0.96
 & 0.20 & 0.98
 & 0.20 & 0.98
 & 0.20 & 0.99 \\
\textbf{MRR@N} 
 & 0.163 & 0.472
 & 0.168 & 0.476
 & 0.168 & 0.476
 & 0.168 & 0.477 \\
\textbf{nDCG@N} 
 & 0.168 & 0.599
 & 0.176 & 0.607
 & 0.176 & 0.607
 & 0.176 & 0.610 \\
\bottomrule
\label{tab:robustness_alpha}
\end{tabular}
\end{table}

\subsection{\texttt{STELLAR} Assertion Generation Quality}
\label{sva_quality}

Having seen the effectiveness of \texttt{STELLAR}'s retrieval mechanism, we now evaluate its impact on the SVA generation.
We compare the performance of \texttt{STELLAR} 
(using top-3 and top-5 retrieved examples, denoted by \textit{K}) 
against a zero-shot baseline\footnote{Although AssertionForge~\cite{bai2025assertionforge} is the closest prior work, the necessary materials for comparison were not available at the time of submission.}, where the LLM generates SVAs on RTL without any context provided.

\textbf{Syntax Correctness.} As shown in Table~\ref{tab:end_to_end_quality}, \texttt{STELLAR} consistently improves the rate of syntactically valid assertions across all models. 
For instance, with \texttt{GPT-4o mini}, \texttt{STELLAR} (\textit{K}=5) improves the syntax correctness rate from 84.9\% to 99.8\%. The results also highlight the benefit of domain adaptation, as \texttt{CodeLlama-7b} consistently outperforms the general-purpose \texttt{Llama-3.1-8b}.
A noticeable fact is that 7B-parameter \texttt{CodeLlama}, when augmented with \texttt{STELLAR}, achieves a syntax correctness at the same level as the much larger proprietary \texttt{GPT-4o mini} model in a zero-shot setting.
This indicates that a domain-specific model combined with a high-precision, structure-aware RAG system like \texttt{STELLAR} can match \texttt{GPT-4o mini} performance in SVA generation tasks.

\textbf{Lexical and Semantic Similarity.} The results in Table~\ref{tab:end_to_end_quality} illustrate a significant disparity between lexical and semantic similarity. Baseline models achieve high zero-shot BERTscores (0.87–0.90), indicating a solid grasp of the core task. However, they fail at stylistic precision, as evident by their low BLEU scores (as low as 0.255). 
\texttt{STELLAR} is critical for bridging this gap by providing high-quality, structurally relevant context.  
It enhances stylistic consistency and 
boosts BLEU scores over \textbf{2.5$\times$} for open-source models. 
It also refines semantic understanding, elevating BERTScores to 0.957 for \texttt{GPT-4o mini}. 
In essence, \texttt{STELLAR} transforms semantically plausible but stylistically flawed outputs into correct and consistent assertions that are useful in a verification environment.

\textbf{Execution-Path Coverage.} 
We evaluate whether the generated assertions cover all required paths for each RTL block. 
An RTL block is counted as covered if the generated SVA set includes at least one assertion for every execution path.
The coverage is computed separately for each design complexity group (\textit{e.g.,} all 2-path designs, all 3-path designs, etc.) 
to obtain a weighted average proportional to each group’s size.
As expected, zero-shot baseline LLMs are highly inconsistent, providing execution-path coverage as low as 29\% (Table~\ref{tab:end_to_end_quality}). 
In contrast, \texttt{STELLAR}'s CRITICAL INSTRUCTION 
guides \texttt{GPT-4o mini} to enhance the execution-path coverage with 100\% accuracy and improves \texttt{Llama-3.1}’s score from 29.5\% to 
93.0\%.

\subsection{Case Studies}

We evaluate STELLAR on three real-world designs, UART, I\textsuperscript{2}C Master, and SHA-3 (Keccak), using \texttt{GPT-4o mini} with \textit{K}=3 retrieved examples. For each design, the Verilog source was pre-processed to extract block-level modules and essential context (Sec.~\ref{sec:STELLAR}) prior to SVA generation. STELLAR is compared to the zero-shot baseline in terms of SVA generation count, syntax-pass rate, and formal coverage. All syntactically valid assertions are subjected to \textit{Cadence Jasper} for 
formal coverage analysis.

As shown in Table~
\ref{tab:case_studies}, for \textbf{UART}, \texttt{STELLAR} substantially improves syntax-pass rate to 89.7\% and produces 3$\times$ more formally proven SVA  than the baseline.
In terms of coverage, \texttt{STELLAR} achieves about 20\% more coverage across multiple formal metrics, demonstrating its effectiveness on deeper formal exploration.
Likewise, for \textbf{I\textsuperscript{2}C master controller~\cite{yan2025assertllm}},
\texttt{STELLAR} outperforms the baseline by generating 125 SVAs and achieving a 95.2\% syntax-pass rate.
Notably, its COI formal coverage is more than 2$\times$ higher than the baseline, demonstrating its superior SVA quality and coverage.
Finally, for \textbf{SHA-3 (Keccak),}
\texttt{STELLAR} generated 44 SVAs with an 90.9\% syntax-pass rate, while the baseline only produced 58 SVAs with 67.2\% pass rate. Despite similar count of valid assertions, \texttt{STELLAR} achieves substantially higher coverage across COI, statement, and toggle metrics. Branch coverage is omitted for SHA-3 because its fix-round, 1600-bit architecture makes many branches inherently unreachable, making this metric not representative for SHA-3.

Overall, the results in  Table~\ref{tab:case_studies} confirm the effectiveness of \texttt{STELLAR} in generating high-quality SVAs across both control-centric designs (UART, I\textsuperscript{2}C) and cryptographic modules (SHA-3).

\begin{table}[h]
\centering
\setlength{\tabcolsep}{3pt}
\caption{Case studies comparing \texttt{STELLAR} (\textit{K}=3) to zero-shot baseline for \#SVA, syntax-pass rate, and formal coverage.}
\label{tab:case_studies}
\smallerspacecaption
\smallerspacecaption
\scriptsize
\begin{tabular}{@{}l|cc|cc|cc@{}}
\toprule
\textbf{Metric} &
\multicolumn{2}{c|}{\textbf{UART}} &
\multicolumn{2}{c|}{\textbf{I\textsuperscript{2}C Master}} &
\multicolumn{2}{c}{\textbf{SHA-3 (Keccak)}} \\
\cmidrule(lr){2-3}\cmidrule(lr){4-5}\cmidrule(lr){6-7}
& \textbf{Baseline} & \textbf{STELLAR} 
& \textbf{Baseline} & \textbf{STELLAR}
& \textbf{Baseline} & \textbf{STELLAR} \\
\midrule
\textbf{\# Generated SVAs} 
& 114 & 184 
& 111 & 125 
& 58 & 44 \\
\textbf{Syntax-Pass Rate (\%)} 
& 55.3 & 89.7 
& 75.0 & 95.2
& 67.2 & 90.9 \\
\midrule
\textbf{COI Coverage (\%)} 
& 67.0 & 85.7 
& 37.7 & 81.3
& 49.5 & 78.3 \\
\textbf{Statement Coverage (\%)} 
& 35.9 & 64.1 
& 34.7 & 85.9
& 40.4 & 78.9 \\
\textbf{Branch Coverage (\%)} 
& 39.1 & 71.7 
& 56.4 & 80.7
& -- & -- \\
\textbf{Toggle Coverage (\%)} 
& 76.0 & 90.0 
& 30.9 & 77.8
& 50.8 & 80.4 \\
\bottomrule
\end{tabular}
\end{table}

\section{Conclusion}
\label{sec:Conclusion}

This paper introduced \texttt{STELLAR}, the first structure-aware framework for automating SVA generation, achieved by guiding LLMs with structurally relevant context retrieved from prior 
codebases. 
Unlike existing works that generate assertions 
from scratch, \texttt{STELLAR} embeds RTL blocks as AST-based structural fingerprints and retrieves similar pairs (RTL, SVA) from the vector embedding space.
Retrieved examples are 
incorporated into a structure-guided prompt that include execution-path instructions, ensuring proper syntax and consistent assertion coverage. 
The method works for fresh designs, which can be added to the knowledge base without retraining, and newly generated, verified SVAs further expand the retrieval knowledge base, enabling scalability for industry.
Experiments on VERT dataset, using proprietary (\texttt{GPT-4o mini}), open-source (\texttt{Llama-3.1-8b}), and domain-adapted (\texttt{CodeLlama-7b}) LLMs yielded significant improvements in syntax, style, and functional correctness. 
The domain-adapted open-source model, paired 
with \texttt{STELLAR}'s retrieval, achieves syntax success rates comparable to the zero-shot proprietary baseline. 
Our results demonstrate the power of structural-guided SVA generation in industrial formal verification.
\begin{acks}
This project is partially supported by NSF grant \#2453861.
\end{acks}

\bibliographystyle{ACM-Reference-Format}
\bibliography{main}

\end{document}